\documentclass[structabstract]{aa}
\usepackage{natbib}
\usepackage{graphicx}
\usepackage{txfonts}
\begin{document}
   \title{Theoretical seismic properties of pre-main sequence $\gamma$~Doradus pulsators}

   \author{M.-P. Bouabid
          \inst{1,2}
          \and
          J. Montalb\'an\inst{2}
	  \and
	  A. Miglio\inst{2,3}
	  \and
	  M.-A. Dupret\inst{2}
	  \and
          A.~Grigahc\`ene\inst{4}
          \and
	  A. Noels\inst{2}
	}

   \institute{UMR 6525 H. Fizeau, UNS, CNRS, OCA, Campus Valrose, 06108 Nice Cedex 2, France\\
              \email{bouabid@oca.eu}
	 \and
              Institut d'Astrophysique et de G\'eophysique de l'Universit\'e de Li\`ege, All\'ee du 6 Ao\^ut, 17 4000 Li\`ege, Belgium\\
         \and
	      School of Physics and Astronomy, University of Birmingham, Edgbaston, Birmingham B15 2TT, United Kingdom\\
	 \and
              Centro de Astrofisica da Universidade do Porto, Rua das Estrelas, 4150-762 Porto, Portugal\\
   }
	
   \date{Received ; accepted }

  \abstract
   {The late A and F-type $\gamma$~Doradus ($\gamma$~Dor) stars pulsate with high-order gravity modes ($g$-modes). The existence of different evolutionary phases crossing the $\gamma$~Dor instability strip raises the question whether pre-main sequence (PMS) $\gamma$~Dor stars exist.}
   {We intend to study the differences between the asteroseismic behaviour of PMS and main sequence (MS) $\gamma$~Dor pulsators as predicted by the current theory of stellar evolution and stability.}
   {We explore the adiabatic and non-adiabatic properties of high-order $g$-modes in a grid of PMS and MS models covering the mass range $1.2 M_\odot < M_* < 2.5 M_\odot$.}
   {We have derived the theoretical instability strip (IS) for the PMS $\gamma$~Dor pulsators. This IS covers the same effective temperature range as the MS $\gamma$~Dor one. Nevertheless, the frequency domain of unstable modes in PMS models with a fully radiative core is greater than in MS models, even if they present the same number of unstable modes. Moreover, the differences between MS and PMS internal structures are reflected in the average values of the period spacing, as well as in the dependence of the period spacing on the radial order of the modes, opening the window to determination of the evolutionary phase of $\gamma$~Dor stars from their pulsation spectra.}
   {}


   \titlerunning{Theoretical seismic properties of PMS $\gamma$~Dor pulsators}
   \maketitle
%

\section{Introduction}
\label{introduction}

The $\gamma$~Dor stars are variable late A and F-type stars. Their variability was identified as caused by pulsation by \cite{balona1994}, and \cite{kaye1999} classified them as a new class of pulsators and defined the features of this group. These stars are pulsating with high-order $g$-modes in a range of periods between  approximately $0.3$ and $3$ days.

The  observational $\gamma$~Dor IS covers a part of the Hertzsprung-Russel Diagram (HRD) between $7200-7700~K$ on the zero-age main sequence (ZAMS) and $6900-7500~K$ above it (\citealt{handler1999}), between 
the solar-like stars domain and the $\delta$~Scuti ($\delta$~Sct) IS. We note that they are located between stars with a deep convective envelope (CE)
and stars with a radiative envelope, in the region of the HR diagram where the depth of the CE changes rapidly with the effective temperature of the star.
Nowadays 66 stars have been confirmed as $bona~fide$ $\gamma$~Doradus (\citealt{henry2007}), and thanks to the space missions  CoRoT (\citealt{baglin2006}) and Kepler (\citealt{gilliland2010}), the number of $\gamma$~Dor candidates is rapidly increasing (see $e.g.$ \citealt{uytterhoeven2008}, \citealt{mathias2009}, \citealt{hareter2010}, \citealt{balona2011}).
The observational limits of the $\gamma$~Doradus IS have been established last by \cite{handler2002} (HS02 hereafter), and in the rest of the paper these limits will be adopted to define the observational $\gamma$~Dor IS.

\begin{figure}[!ht]
\centering
\includegraphics[width=70mm,height=60mm]{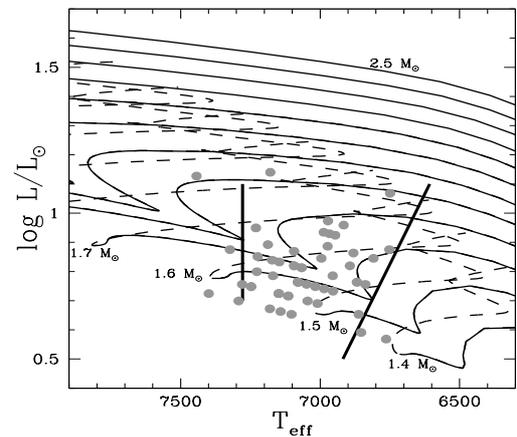}
\caption{PMS and MS (respectively full $\&$ dashed thin lines) evolutionary tracks that cross the observational $\gamma$~Dor IS (thick lines - HS02). Grey points represent the $bonafide$ $\gamma$~Dor stars from \cite{henry2007}.}
\label{HRdiag}
\end{figure}

\cite{guzik2000} use the frozen convection approximation to propose the modulation of the radiative flux at the base of the CE as the excitation mechanism. This mechanism has been confirmed by \cite{dupret2005} using a time-dependent convection (TDC) treatment (\citealt{gabriel1996}, \citealt{grigahcene2005}).
Because the depth of the CE plays a major role in the driving mechanism of $\gamma$~Dor pulsations, the theoretical predictions of stability are very sensitive to the parameter $\alpha$ defining the mean free path of a convective element ($\Lambda=\alpha \times H_p$, where $H_p$ is the pressure scale height) in the classical mixing-length treatment of
convection (MLT, \citealt{bohmvitense1958}). Using TDC treatment, \cite{dupret2004} obtained good agreement between theoretical and observational  $\gamma$~Dor IS for models computed with $\alpha = 2.00$.

   These theoretical works on $\gamma$~Dor stars systematically studied MS models. However, as shown in Fig.~\ref{HRdiag}, MS and PMS\footnote{We consider as PMS models those before the onset of the central H-burning \textit{at equilibrium}.} evolutionary tracks cross the observed 
   $\gamma$~Dor IS. Although the time spent by a star of 1.8~$M_\odot$ to cross the IS during its PMS evolution is ten times less than the time spent during the MS phase in the same region of HR diagram, recent photometric observations of young clusters (NGC~884, \citealt{saesen2010}) have revealed the presence of $\gamma$~Dor candidates  that, given the age of the cluster, should be in the PMS phase. A strong effort has also been made to find PMS pulsators in NGC2264 (\citealt{zwintz2009}), another young open cluster. Moreover, the PMS/MS status of HR~8799, a $\gamma$~Dor variable hosting four planets (or brown dwarfs) (\citealt{marois2008}, \citealt{marois2010}), is still a matter of debate (\citealt{moya2010a}, \citealt{moromartin2010}).

It is then timely to theoretically study the seismic properties of PMS $\gamma$~Dor in order to derive possible differences between their spectra and those of $\gamma$~ Dor in the MS phase.
This is the aim of this paper, which is structured as follows. In Sect.~\ref{internalstructure} we compare the internal structure of MS and PMS models computed as described in Sect.~\ref{stellarmodels}. The effects of these structure differences on the properties of the adiabatic frequency spectra of PMS and MS $\gamma$~Dor are analysed in Sect.~\ref{adiabaticstudy}.
In Sect.~\ref{nonadiabaticstudy}, we focus on the excitation and damping of $\gamma$~Dor pulsations in PMS models and on the differences between PMS and MS non-adiabatic quantities. A summary is finally given in Sect.~\ref{conclusion}.


\section{Stellar models}
\label{stellarmodels}

Stellar models were computed with the stellar evolutionary code CLES (\citealt{scuflaire2008a}) for masses between $1.2$ and $2.5 M_\odot$, initial helium mass fraction $Y_0=0.28$, and initial metal mass fraction $Z_0=0.02$. We adopted the AGS05 \citep{asplund2005} metal mixture, and the corresponding opacity tables were computed with OP\footnote{http://opacities.osc.edu/} (\citealt{badnell2005}) facilities  and completed at low temperature ($\log T < 4.1$) with  \cite{ferguson2005} opacity tables. We used the OPAL2001 equation of state (\citealt{rogers2002}) and the nuclear reaction rates from NACRE compilation (\citealt{angulo1999}), except for the $^{14}$N$(p,\gamma)^{15}$O nuclear reaction, for which we adopted the cross section from \cite{formicola2004}. Surface boundary conditions at $T=T_{\rm eff}$ were provided by ATLAS model atmospheres (\citealt{kurucz1998}).
Convection was treated using the MLT formalism (\citealt{bohmvitense1958}) with an MLT parameter $\alpha = 2.00$. We considered models without and with convective core overshooting,  with an overshooting parameter $\alpha_{\rm ov} =0.20$ and the distance of extra-mixing given by $d_{\rm ov}=\alpha_{\rm ov} \times \min(r_{cc},H_p)$, where $r_{cc}$ is the convective core radius. 


\section{Internal structure of PMS and MS models}
\label{internalstructure}

\begin{figure}[!ht]
\centering
\includegraphics[width=70mm,height=60mm]{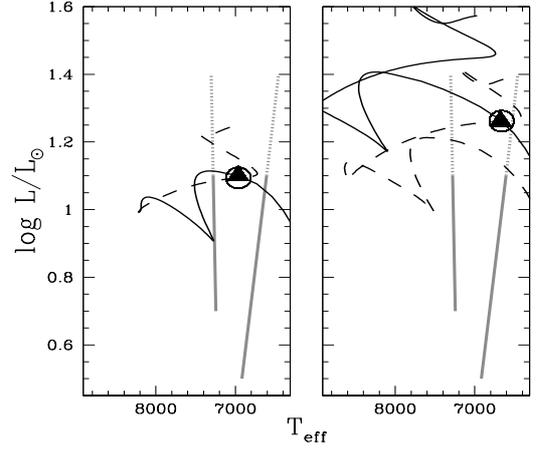}
\caption{Evolutionary tracks for different models crossing the $\gamma$~Dor observational IS (thick grey lines - HS02). Left panel: $1.8 M_\odot$ evolutionary track whose PMS (full line) and MS (dashed line) phases intersect inside the IS (circle: PMS model - triangle: MS model). Right panel: evolutionary tracks for  $1.9$ and $2.1 M_\odot$ showing the same HR location for a model of $2.1 M_\odot$ in the PMS phase (circle) and a model of $1.9 M_\odot$ at the end of the MS phase (triangle).}
\label{HRcompare}
\end{figure}

During the Hayashi track evolution, the global contraction of the star leads to the  increase in density and temperature. This tends to decrease the opacity and a radiative core develops and grows in mass as the star evolves. When the central temperature is high enough ($\sim1.7~10^7$~K), the nuclear CN subcycle starts.
As the $^{12}$C$(p,\gamma)^{13}$N$(\beta^+\nu)^{13}$C$(p,\gamma)^{14}$N nuclear reaction is very sensitive to the temperature ($\propto T^{19}$), a convective core (CC) appears in the star. 
The mass fraction of this CC changes as the star evolves toward the ZAMS and for stars slightly more massive than the Sun, this CC remains during the MS.

We see in Fig.~\ref{HRdiag} that the evolutionary phase at which a PMS star crosses the $\gamma$~Dor IS depends on its mass. While lower mass stars have already developed a CC, the more massive stars still have a contracting, chemically homogeneous radiative core when they cross the IS. We recall that the PMS phase is a brief evolutionary phase compared to the MS one, especially if we only consider the PMS phase before the onset of nuclear reactions.

The properties of high-order $g$-mode spectrum are determined by the matter stratification in the star, 
which is described by the Brunt-V\"ais\"al\"a frequency $N$
\begin{equation}
\label{bvfrequency}
\centering
N^2=g\left(\frac{1}{\Gamma_1}\frac{\textrm{d} \ln P}{\textrm{d}r}-\frac{\textrm{d} \ln \rho}{\textrm{d}r}\right) \propto \left(\nabla_{\rm ad}-\nabla+\frac{\varphi}{\delta}\nabla_{\mu}\right)
\end{equation}
with $g$ the local gravity, $\rho$ the local density, $P$ the local pressure, $r$ the local radius, $\Gamma_1$ the first adiabatic exponent, $\nabla_{\rm ad}$ and $\nabla$ are respectively the adiabatic and stellar temperature gradients, $\nabla_{\mu}$ the mean molecular weight gradient, and $\varphi$ and $\delta$ are the partial derivatives of density with respect to $\mu$ and temperature, respectively.
It is possible to highlight the direct link between the $g$-mode spectrum and the internal structure of the star from the first-order asymptotic theory (\citealt{tassoul1980}). In this approximation, the period of a high-order $g$-mode of degree $\ell$ in a star with a CC and a CE is written as
\begin{equation}
\label{asymptoticperiod}
\centering
	P_k = \frac{\pi^2}{\sqrt{\ell(\ell+1)} \int_{r_0}^{r_1}{\frac{N}{r} \textrm{d}r}} \left( 2k + 1 \right),
\end{equation}
with $k$ the mode radial order, $r_0$ and $r_1$ the limits of the $g$-modes cavity defined by $\omega^2 \ll N^2, S_\ell^2$, where $\omega$ is the mode frequency and $S_\ell$ the Lamb frequency for $\ell$-degree modes.

Eq.~(\ref{asymptoticperiod}) makes the role played by $N$ in the central regions evident in the determination of $\gamma$~Dor oscillation properties.
Because of the differences between PMS and MS stellar structures, we expect differences between the seismic properties of stars in these evolutionary phases.

Fig.~\ref{HRcompare} presents the evolutionary tracks of $1.8 M_\odot$ (left panel), 1.9, and $2.1 M_\odot$ (right panel) models. PMS and MS $1.8~M_\odot$ evolutionary tracks cross each other at different points of the HRD, in particular inside the $\gamma$~Dor IS.
The comparison of two models at the same location in the HRD allows us to eliminate the effects of different effective temperatures and radii on the stellar structure.

\begin{figure}[!ht]
\centering
\includegraphics[width=70mm,height=60mm]{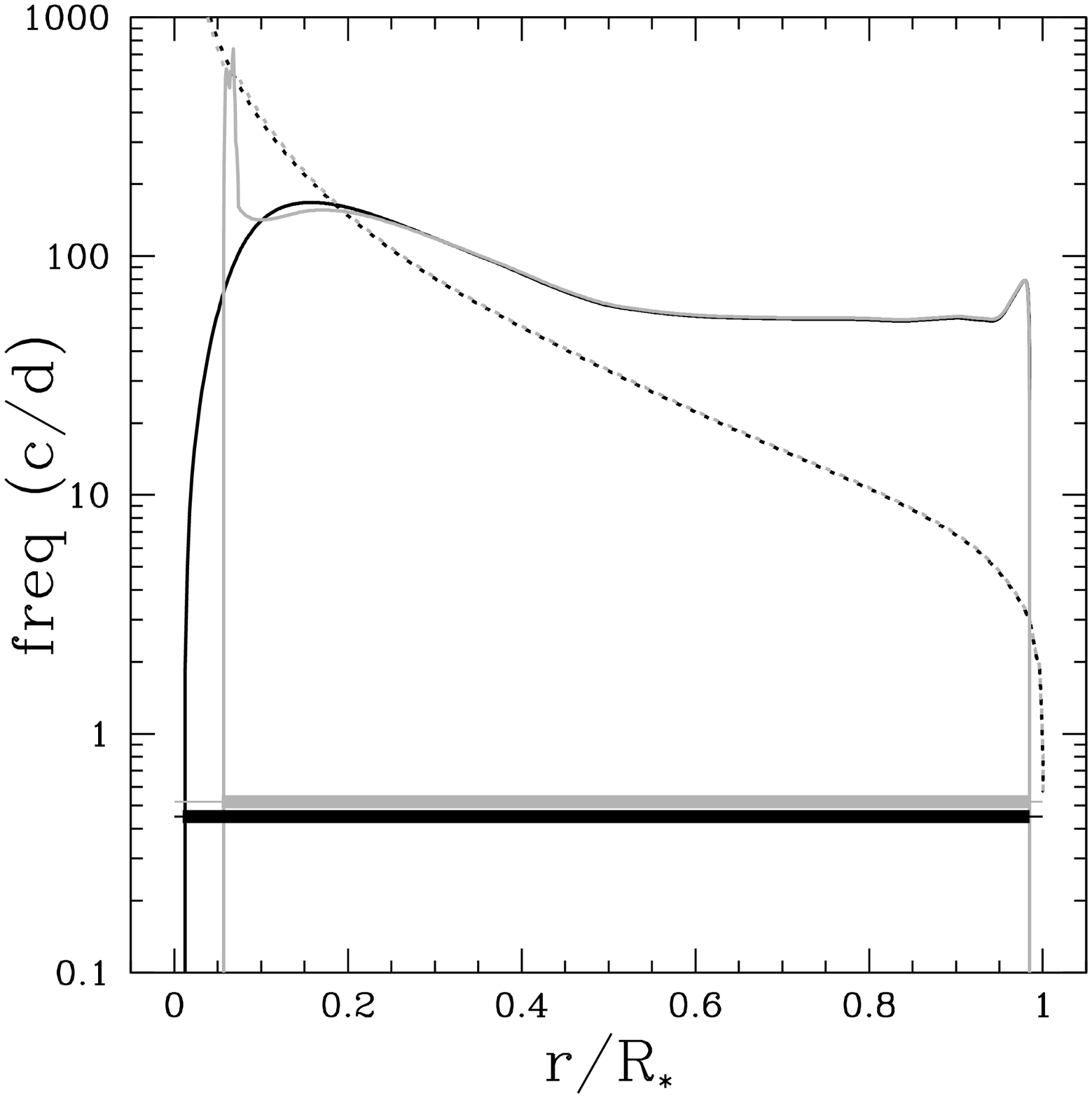}
\includegraphics[width=70mm,height=60mm]{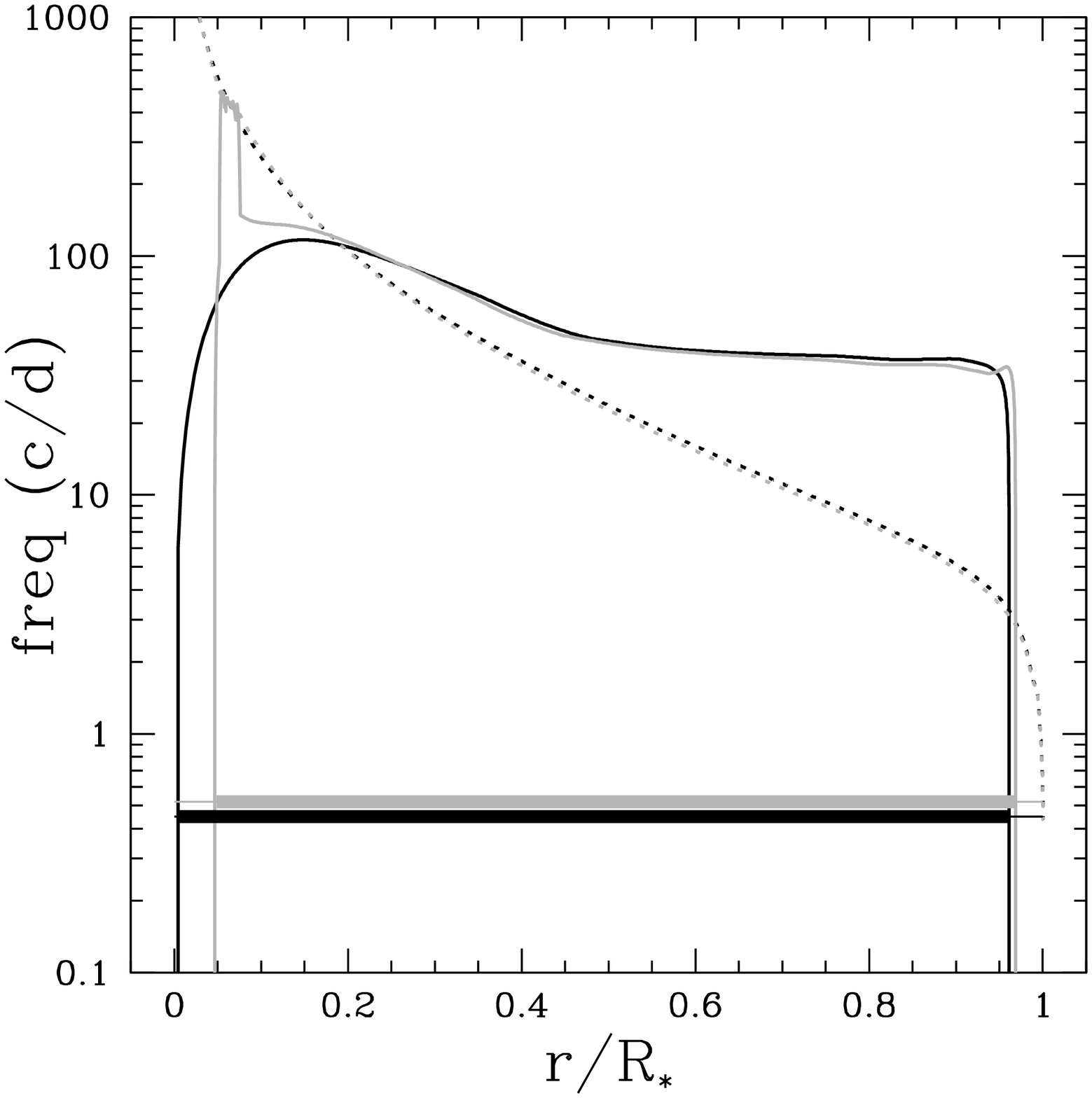}
\caption{Propagation diagram. PMS (black), MS (grey) $N$ (full lines), and $S_{\ell = 1}$ (dotted lines) frequencies for models with the same mass (top panel) and for models with different masses (bottom panel) presented in Fig.~\ref{HRcompare}. The thick horizontal lines represent the propagation zone for typical $\gamma$~Dor $g$-modes.}
\label{BVfreq}
\end{figure}

In Fig.~\ref{BVfreq} (top panel), we plot the $N$ profiles of the two $1.8 M_\odot$ models.
Because of the same mean density and $T_{\rm eff}$, PMS and MS Brunt-V\"ais\"al\"a frequency profiles show similar behaviour in the outer layers, and the bottom of the CE is located at the same depth in these two models. Both $N$ profiles also present a bump in the inner layers, which is due to the density distribution.
In the PMS model, the onset of the $^{12}$C$(p,\gamma)^{13}$N$(\beta^+\nu)^{13}$C$(p,\gamma)^{14}$N nuclear reaction has already led to the development of a small CC. The MS model, in addition to a larger CC,  presents a sharp  variation in the Brunt-V\"ais\"al\"a frequency profile due to the  mean molecular weight gradient let by the CC during the MS evolution.  The $N$ profile of a PMS star is smooth, even if a small CC already exists.

As above mentioned, more massive PMS models crossing the $\gamma$~Dor IS are in a less evolved phase than lower mass ones. In order to have MS models with low enough effective temperature and
high enough luminosity to meet more massive PMS  tracks, we have to consider models with CC overshooting ($\alpha_{\rm ov}~=~0.20$). 
The intersection  between $2.1 M_\odot$ and $1.9 M_\odot$  evolutionary tracks in the $\gamma$~Dor IS occurs when  the PMS $2.1 M_\odot$ model has a 
radiative and quasi chemically homogeneous structure while the $1.9 M_\odot$ one is  at the end of its MS (Fig.~\ref{HRcompare} - right).
Compared with the 1.8~$M_\odot$ models described in the previous paragraph, the 2.1~$M_\odot$ PMS model is  less evolved and still has not developed a CC; on the other hand, the 1.9~$M_\odot$ MS model is more evolved and presents a larger $\nabla_\mu$. Moreover, although these models have the same radius, they do not have the same gravity, so the density profiles in the outer layers and the depth of the convective envelope are different (Fig.~\ref{BVfreq} - bottom panel)

\section{Adiabatic analysis - Period spacing}
\label{adiabaticstudy}

PMS and MS structures in the $\gamma$~Dor region differ particularly in their 
central layers. We performed an adiabatic oscillation study of our grid of stellar models in order to link these differences to asteroseismic quantities. The adiabatic oscillation frequencies were computed with the code LOSC (\citealt{scuflaire2008b}).

\subsection{Evolution of the period spacing in the first order asymptotic approximation}
\label{periodspacingevolution}

\begin{figure}[!ht]
\centering
\includegraphics[width=70mm,height=60mm]{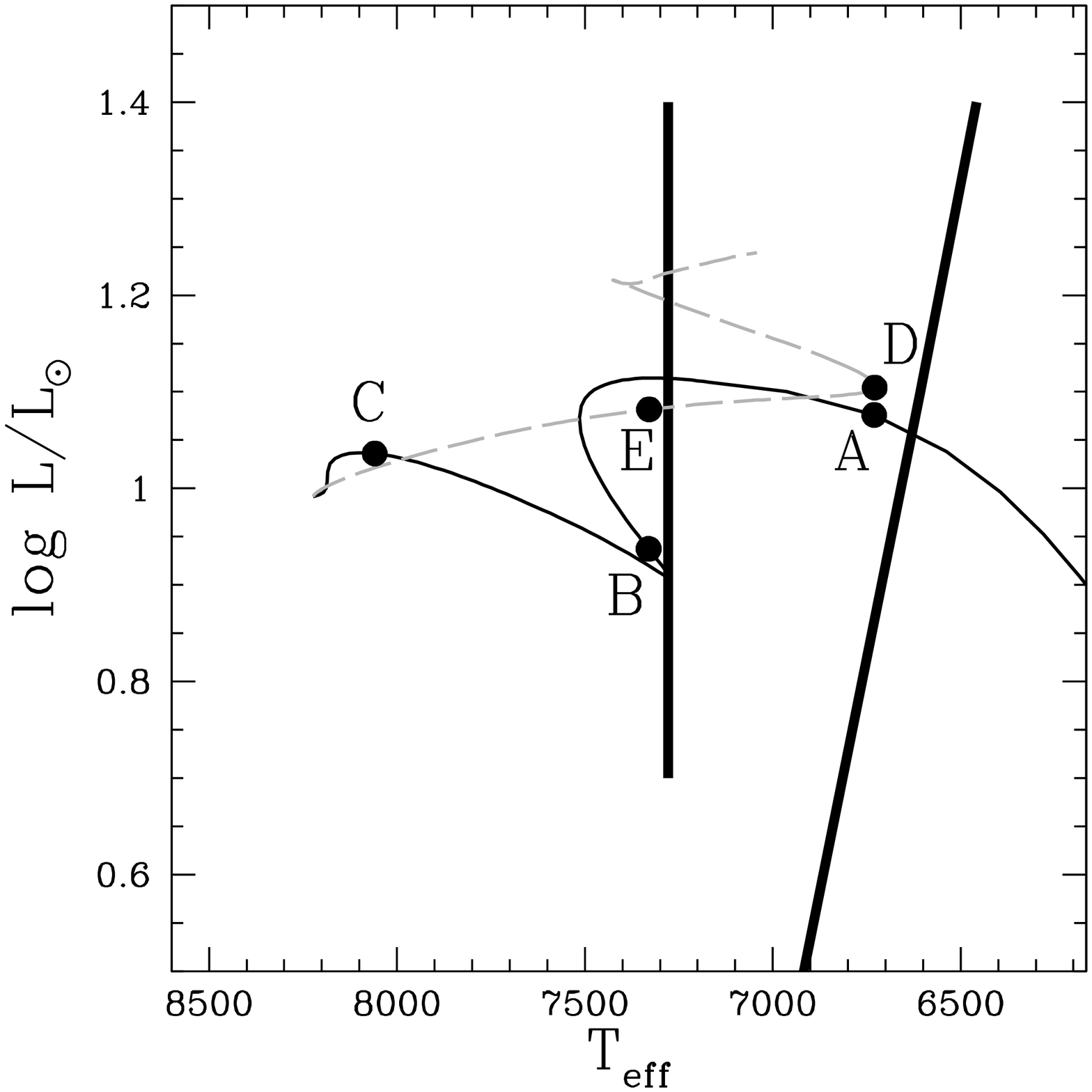}
\includegraphics[width=70mm,height=60mm]{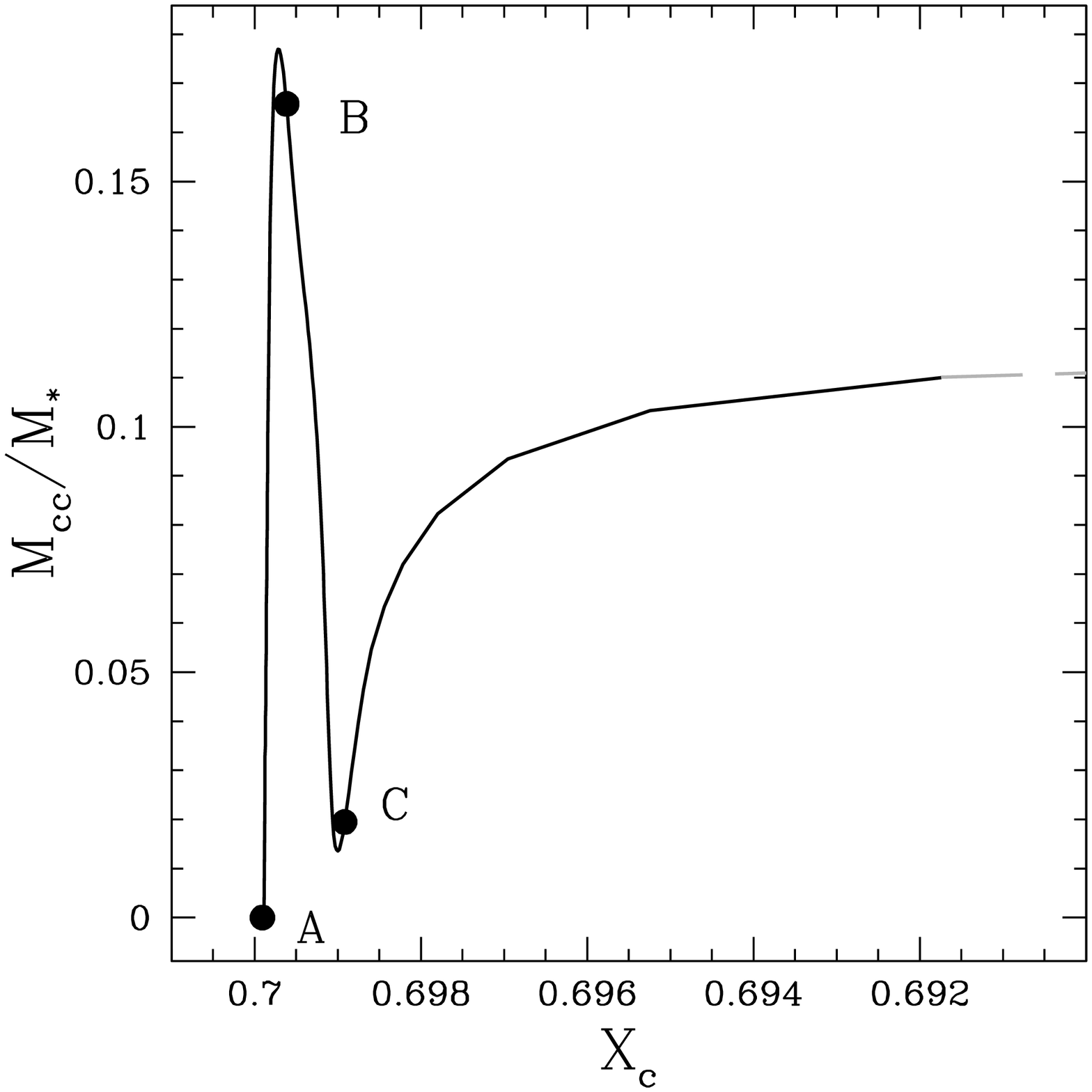}
\includegraphics[width=70mm,height=60mm]{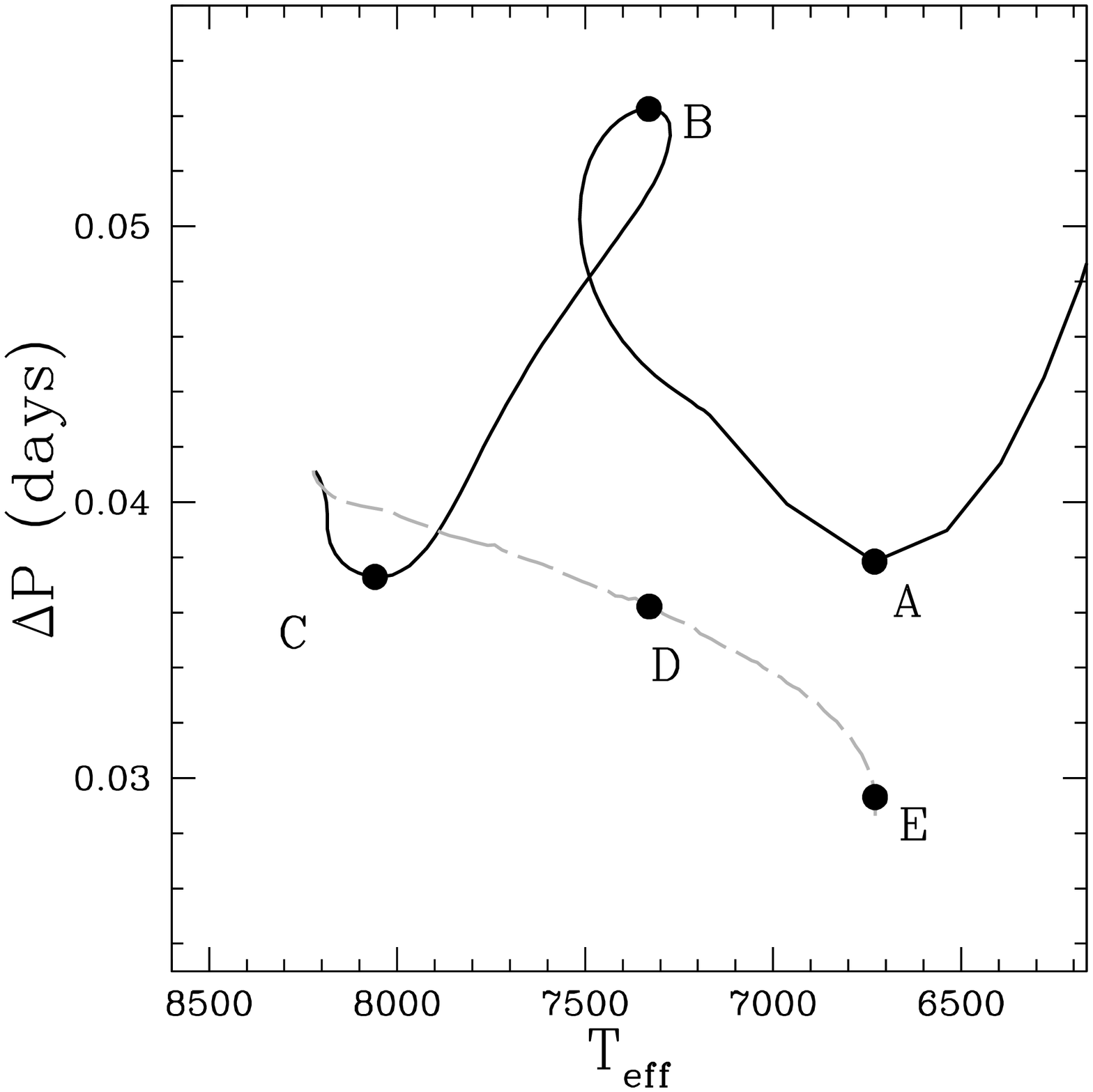}
\caption{Top panel: Evolution of a $1.8 M_\odot$ star in the HRD (PMS phase in full black and MS phase in dashed grey) crossing the $\gamma$~Dor IS (thick black lines - HS02).
Middle panel: Variation in its CC mass from the PMS phase to the early MS phase (${\rm X_c} = 0.69$).
Bottom panel: Evolution of the $\ell = 1$ modes period spacing as a function of the effective temperature of the star from the PMS to the end of MS (point E).
}
\label{periodspaceall}
\end{figure}

From Eq.~(\ref{asymptoticperiod}) we obtain the expression for the asymptotic period spacing between two $g$-modes with consecutive radial orders and the same $\ell$-degree
   \begin{equation}
	\label{meanperiodspacing}
	\centering
	\Delta P = P_{k+1} - P_{k} = \frac{2 \pi^2}{\sqrt{\ell(\ell+1)}\int_{r_0}^{r_1}{\frac{N}{r}dr}}.
   \end{equation}
This asymptotic approximation predicts a constant value of the period spacing, independent  of the radial order $k$.

	Fig.~\ref{periodspaceall} (bottom panel) shows the evolution of the asymptotic period spacing of a $1.8 M_\odot$ star from the PMS to the end of the  main sequence. The variation in the period spacing value during the stellar evolution seems strongly linked to the evolution of the CC (middle panel). The largest difference between PMS and MS period spacing at a given effective temperature is about $0.02$ days and corresponds to the maximum mass fraction of the CC during the PMS phase (Fig.~\ref{periodspaceall} - point $B$ compared to point $D$).

\begin{figure}[!ht]
\centering
\includegraphics[width=70mm,height=60mm]{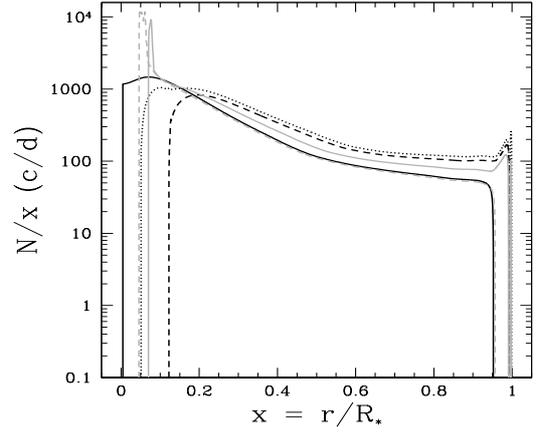}
\caption{$\frac{N}{x}$ ratio as a function of the normalized radius for models A, B, C (respectively full, dashed, dotted black lines), D and E (respectively full and dashed grey lines).}
\label{freqbvperiodspace}
\end{figure}

In Fig.~\ref{freqbvperiodspace}, we present the $\frac{N}{x}$ ratio for models A to E (with $x$ the normalized radius). For high-order $g$-modes in $\gamma$~Dor pulsators, the limits of the propagation cavity almost coincide with the boundaries  of convective regions, and the contribution of $\frac{N}{x}$ to the $\Delta P$ value comes  from the whole radiative zone. Because of its large CC, model B has a low $\int{\frac{N}{x}dx}$ value, leading to a high value of $\Delta P$ (see Fig.~\ref{periodspaceall}). 
Even if models A, C, and D have very different structures, hence different $N$ profiles, the values of the integral over the propagation regions lead to similar values of $\Delta P$. The high contribution to $\int{\frac{N}{x}dx}$ from the $\nabla_\mu$ region in model D is within an order of magnitude of the contribution of the most central layers of model A. On the other hand, the higher value of $\frac{N}{x}$ in $\nabla_\mu$ region is offset by the lower value in the outer layers when model D is compared to model C.

\begin{figure}[!ht]
\centering
\includegraphics[width=70mm,height=60mm]{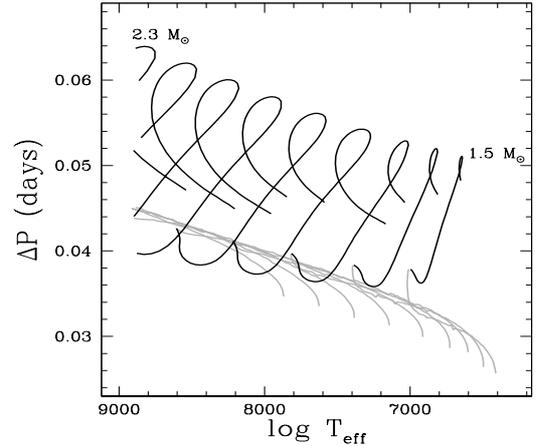}
\caption{Evolution of the $\ell = 1$ modes period spacing as a function of the effective temperature of PMS and MS models for evolutionary tracks between $1.5 M_\odot$ and $2.3 M_\odot$ in the $T_{\rm eff}$ range of the $\gamma$~Dor IS.}
\label{periodspacemasses}
\end{figure}

Fig.~\ref{periodspacemasses} illustrates the evolution of the $\ell = 1$ period spacing for models between 1.5 and $2.3 M_\odot$. Although the value of period spacing for models in the  PMS phase is generally higher than for models in the MS phase, they approach close to the MS. As a consequence, we conclude that the average value of $\Delta P$ cannot be used alone to distinguish PMS $\gamma$~Dor from  MS ones.

\subsection{Oscillatory signature in the period spacing}
\label{oscillatorysignature}

\cite{miglio2008} investigated the properties of high-order $g$-modes in MS stellar models and show that the sharp variation in the $N$ profile at the limit of the CC in an MS model lets clear asteroseismic signature: the oscillation of the period spacing around its mean value.
They define $\delta P_k$ as the difference between the periods of a star showing such a sharp variation in $N$ and the periods of an otherwise fictitious smooth model with the same value of $\int_{r_0}^{r_1}{\frac{N}{r}dr}$. Assuming the \cite{cowling1941}, JWKB (see $e.g.$ \citealt{gough2007}) and asymptotic approximations, they derive the following expression of $\delta P_k$ for a $N$ profile modelled by a step function
\begin{equation}
	\label{periodspacingstructure}
	\delta P_k \propto \frac{\Pi_0}{L}\frac{1-\nu^2}{\nu^2}\cos\left(2 \pi \frac{\Pi_0}{\Pi_\mu} k + \Phi \right)
\end{equation}

\noindent
where $\Phi$ is a phase constant, $\nu = \left(\frac{N_+}{N_-}\right)$ with $N_+$ and $N_-$ respectively the values of the Brunt-V\"ais\"al\"a frequency at the outer and inner borders of the $\mu$-gradient region. The local buoyancy radius is defined as

\begin{equation}
	\label{buoyancylocalradius}
	\Pi_x^{-1} = \int_{x_0}^{x}{\frac{N}{x}dx}
\end{equation}
Then the total buoyancy radius is
\begin{equation}
	\label{buoyancytotalradius}
	\Pi_0^{-1} = \int_{x_0}^{x_1}{\frac{N}{x}dx},
\end{equation}
and the buoyancy radius at the sharp variation is
\begin{equation}
	\label{buoyancysharpvariationradius}
	\Pi_\mu^{-1} = \int_{x_0}^{x_\mu}{\frac{N}{x}dx}
\end{equation}
with $x_1$ the normalized radius at the top of the propagation cavity, and  $x_0$ and $x_\mu$ the normalized radii respectively at the boundary of the convective core and at the location of the sharp variation of the $N$ profile produced by $\nabla_\mu$.

Eq.~(\ref{periodspacingstructure}) describes  the signature of the abrupt change in the $N$ profile as a sinusoidal component in the period spacing variation with an amplitude proportional to the sharpness of the change in $N$ and a periodicity in terms of the radial order $k$ given by
\begin{equation}
	\label{radialorderbuoyancyradius}
	\Delta k \sim \frac{\Pi_\mu}{\Pi_0}.
\end{equation}

\begin{figure}[!ht]
\centering
\includegraphics[width=70mm,height=60mm]{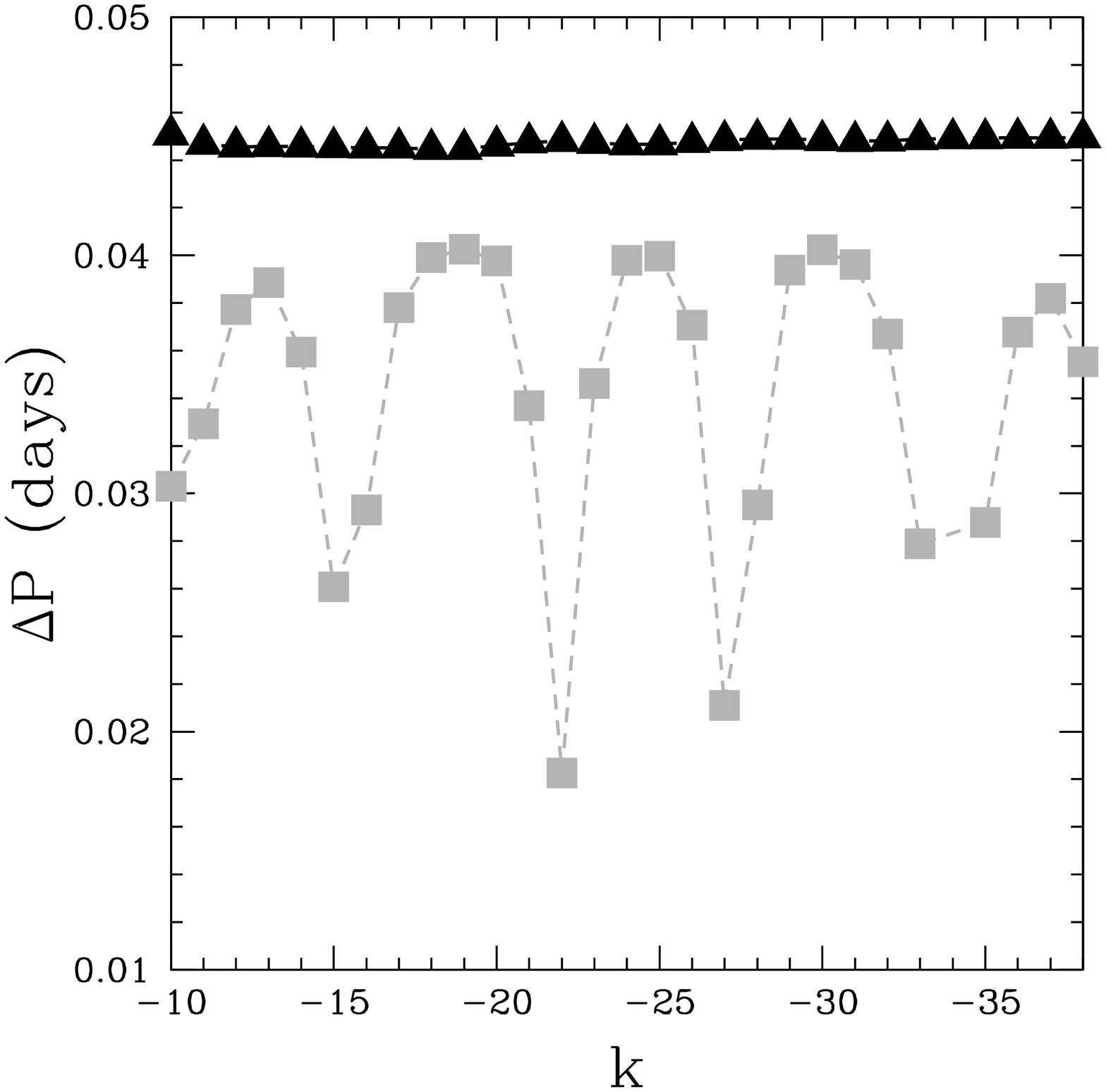}
\includegraphics[width=70mm,height=60mm]{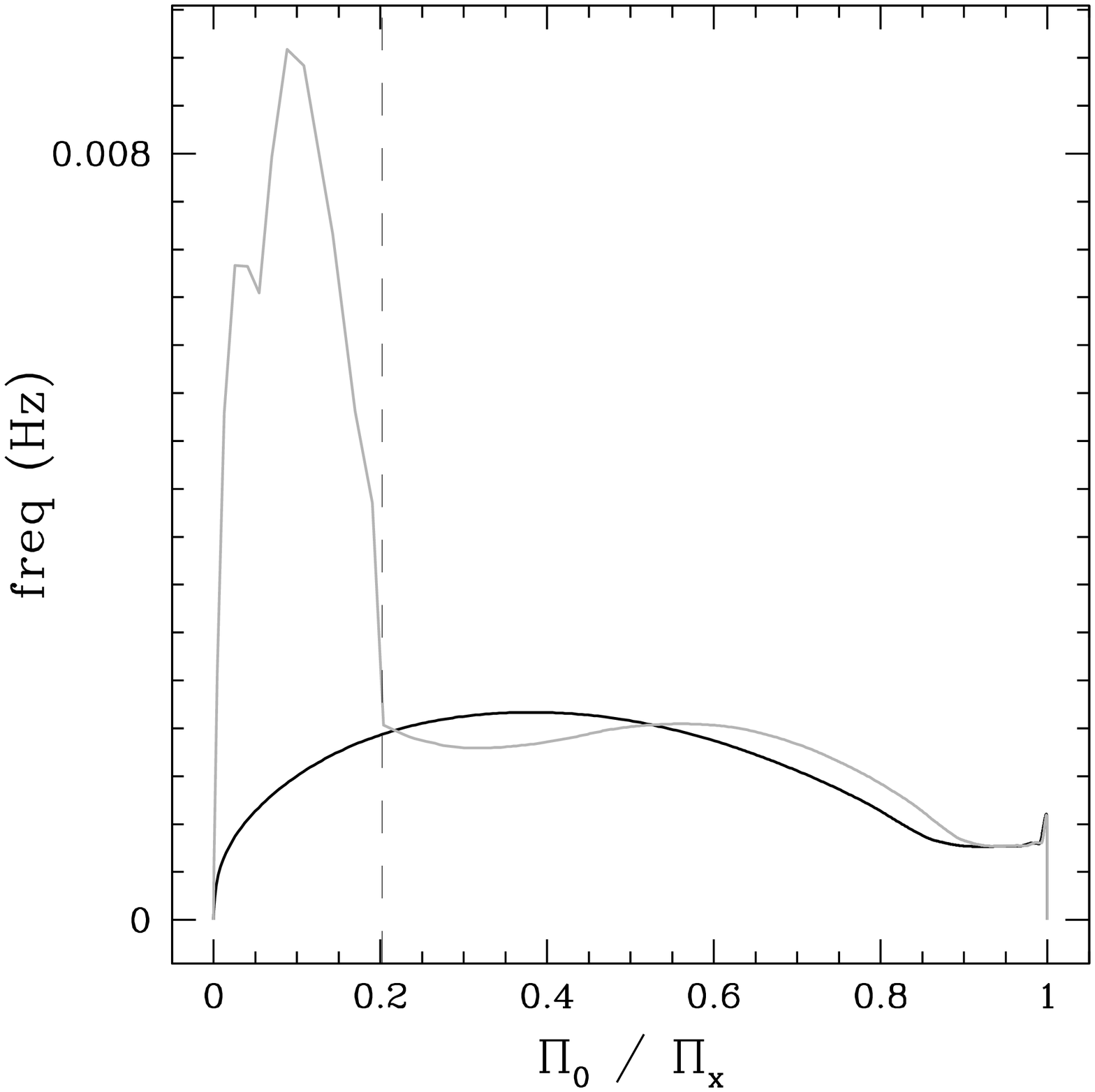}
\caption{Top panel: Period spacing structure for $\ell=1$ modes as a function of the radial order $k$ of the modes for PMS (black triangles) and MS (grey squares) models with the same mass ($M = 1.8 M_\odot$). Bottom panel: The Brunt-V\"ais\"al\"a frequency versus $\Pi_0/\Pi_x$ of these $1.8 M_\odot$ PMS (black) and MS (grey) models. The dashed vertical line indicates the location of $\frac{\Pi_\mu}{\Pi_0}$ for the MS model.}
\label{periodspacestructure}
\end{figure}

Fig.~\ref{periodspacestructure} (top panel) presents the $\ell = 1$ period spacing as a function of the radial order $k$ for the  $1.8 M_\odot$ PMS and MS models described in Sect.~\ref{internalstructure}. The bottom panel presents the $N$ profiles of these two models versus the relative buoyancy radius. In the case of the PMS model, since $\nabla_\mu$ is not significant,  no sharp variation in $N$  appears in the radiative region, making $\Delta P$ almost constant. On the contrary,  the sharp feature in the Brunt-V\"ais\"al\"a frequency profile of the MS model introduces  an oscillatory variation in the period spacing values. For the MS model, we see in Fig.~\ref{periodspacestructure} (bottom panel) that $\left( \frac{\Pi_0}{\Pi_\mu} \right)^{-1}$ value is about 5, which is the mean number of radial orders between two consecutive minima of the period spacing value (Fig~\ref{periodspacestructure} - top panel). This oscillation does not show a sinusoidal behaviour as expected from Eq.~(\ref{periodspacingstructure}) because the approximations used to derive this expression are only valid for small variations relative to the smooth model. Nevertheless, we verify the relation between the oscillation periodicity in terms of $k$ and the buoyancy radius of the sharp variation ($\Pi_\mu$).

Therefore, the deviation of the period-spacing behaviour from the constant value expected from the first-order asymptotic approximation can be used to distinguish   between PMS and MS $\gamma$~Dor pulsators. In fact, the periodicity of the variation of $\Delta P$  is directly linked to the location of the chemical gradient at the  border of the convective core and the amplitude of this variation, and its dependence on $k$  give information on the $N$ profile \citep[see][for details] {miglio2008}.

\section{Non-adiabatic analysis}
\label{nonadiabaticstudy}

This section presents the results of our non-adiabatic oscillation study. We computed $\ell = 1$ and $\ell =2$ modes for our grid of stellar models using the non-adiabatic code MAD with the time-dependent treatment of convection (\citealt{dupret2001}, \citealt{grigahcene2005}).

\begin{figure}[!ht]
\centering
\includegraphics[width=70mm,height=60mm]{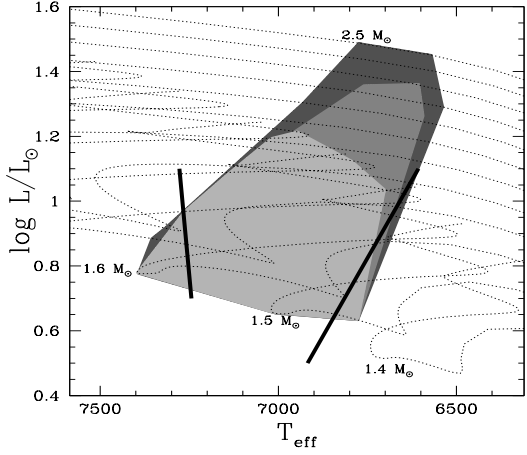}
\caption{Theoretical IS for MS (light grey), MS with overshooting (grey) and PMS (dark grey) $\gamma$~Dor stars compared to the observed $\gamma$~Dor IS (thick black lines - HS02). 
The thin dotted lines represent evolutionary tracks for models between $1.4$ and $2.5 M_\odot$.
}
\label{ISpms}
\end{figure}

In Fig.~\ref{ISpms}, we present the theoretical IS for PMS and MS $g$-mode pulsators in the $\gamma$~Doradus region of the HRD. PMS high-order $g$-modes pulsators are expected in this region, and the unstable PMS models cover the same effective temperature range as the MS $\gamma$~Dor IS. The only difference between both IS lies in the extension of the PMS one toward higher luminosities. With our set of physical parameters (see Sect.~\ref{stellarmodels}), we cannot obtain MS models in the upper region of the HRD where we have young massive PMS pulsators, even with CC overshooting ($\alpha_{\rm ov} = 0.20$).

\begin{figure}[!ht]
\centering
\includegraphics[width=70mm,height=60mm]{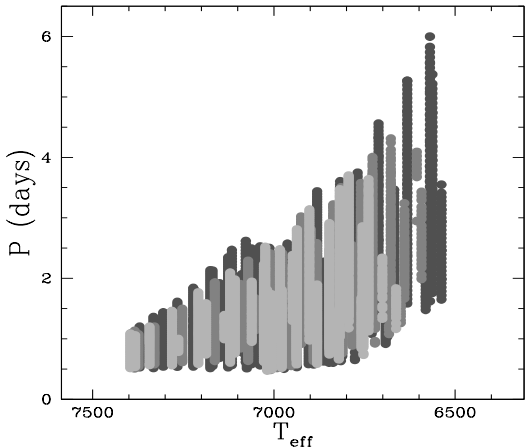}
\caption{Period range of $\ell =1$ unstable modes for all the computed PMS (dark grey), MS (light grey) and MS with overshooting (grey) models as a function of their effective temperature.}
\label{periodteff}
\end{figure}

Fig.~\ref{periodteff} shows the $\ell = 1$ instability domain versus the effective temperature for all PMS and MS models with unstable $g$-modes. The PMS and MS instability domain show a similar behaviour in the full $T_{\rm eff}$ range, apart from a region near $P_k~\sim~6$ days where we have young massive PMS models.
To study this behaviour we  focus on two non-adiabatic quantities
\begin{itemize}
\item the dimensionless work integral ($W$) defined as
\begin{equation}
\label{workintegral}
\centering
	W_m = -\frac{R^{3/2}}{2\sqrt{GM}\omega} \frac{\int_0^m{Im \left(\frac{\delta \rho^*}{\rho} T \delta s \right) \left( \Gamma_3 - 1 \right) \textrm{d}m}}{\int_0^M \left( {|\xi_r|^2 + \ell \left( \ell + 1 \right) |\xi_h|^2} \right) \textrm{d}m}
\end{equation}
with $R$ and $M$ respectively the radius and the mass of the model, $G$ the gravitational constant, $\omega$ the frequency of the mode, $s$ the specific entropy, $\Gamma_3$ the third adiabatic exponent,
$*$ defines the complex conjugate of a quantity, and $\xi_r$ and $\xi_h$ are respectively the radial and the horizontal components of the displacement. The work integral of a given mode gives the contribution of a region of the star to the excitation/damping of this mode. Regions where $W$ increases contribute to drive the mode, while regions where $W$ decreases contribute to its damping. Finally, a mode is unstable if $W > 0$ at the surface of the star.

\item the eigenfunction $\delta P/P$ that is linked to the amplitudes of the oscillations. Only regions where $\delta P/P$ is large enough play a significant role in the driving/damping of the mode.
\end{itemize}

Before getting to the heart of this study, it is important to recall the excitation and damping processes that occur in $\gamma$~Dor stars.
As developed in \cite{guzik2000} and \cite{dupret2005}, the pulsations in $\gamma$~Dor stars come from a periodic modulation of flux at the bottom of the convective envelope, therefore the behaviour of the unstable modes is strongly linked to the depth of the CE. If two models have the same CE depth, they then have the same $\gamma$~Dor modes excitation capacity. Nevertheless, this is not sufficient to obtain the same period ranges of unstable modes. Indeed, another important process is the radiative damping that occurs in the central layers of the star.
\cite{dupret2005} show that $\gamma$~Dor $g$-modes are unstable if and only if the excitation mechanism at the bottom of the convective envelope overcomes the damping mechanism.
In the case of high-order $g$-modes, the eigenfunctions oscillate quickly in the $g$-mode cavity, leading to high values of their second derivatives, which play a role in the work-integral expression. This oscillation leads to a value of the radiative damping that is high enough to be more efficient than the $\gamma$~Dor driving mechanism.
For low-order $g$-modes, the absolute value of the eigenfunctions is significant in 
the $g$-mode cavity compared to its value in the more superficial layers, which implies an efficient radiative damping in the $g$-mode cavity.
Within the asymptotic theory, \cite{vanhoolst1998}, \cite{dziembowski2001}, and \cite{godart2009} propose a simple expression of the radiative damping $\eta$, which is proportional to
\begin{equation}
	\label{radiativedamping}
	\eta \propto \int_{r_0}^{r_1} \frac{\nabla_{ad} - \nabla}{\nabla} \frac{\nabla_{ad} N g L}{P r^5} \textrm{d}r.
\end{equation}
\noindent 
We can rewrite $\eta$ in a dimensionless form as
\begin{equation}
	\label{dimlessradiativedamping}
	\eta \propto \frac{R^6}{(GM)^2} \int_{r_0}^{r_1} \frac{\nabla_{ad} - \nabla}{\nabla} \frac{\nabla_{ad} N g L}{P r^5} \textrm{d}r = \int_{x_0}^{x_1} I(x) \textrm{d}x.
\end{equation}
\noindent 
The radiative damping depends strongly on the matter stratification in the central layers of the star because of the factors $1/r^5$ and $N$.

\begin{figure}[!ht]
\centering
\includegraphics[width=70mm,height=60mm]{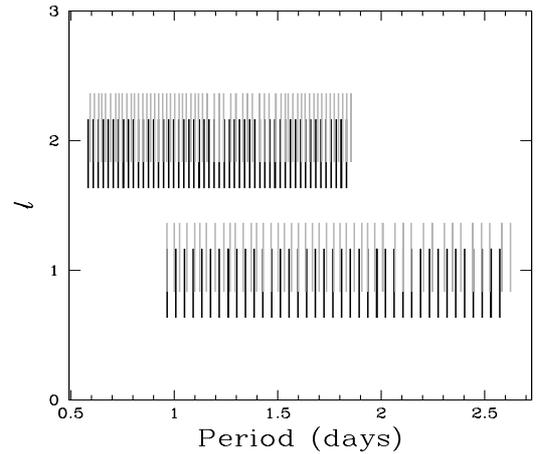}
\caption{Period range for $\ell = 1$ and $\ell = 2$ unstable modes of PMS (black) and MS (grey) $1.8 M_\odot$ models. }
\label{periodsamemass}
\end{figure}

\begin{figure}[!ht]
\centering
\includegraphics[width=70mm,height=70mm]{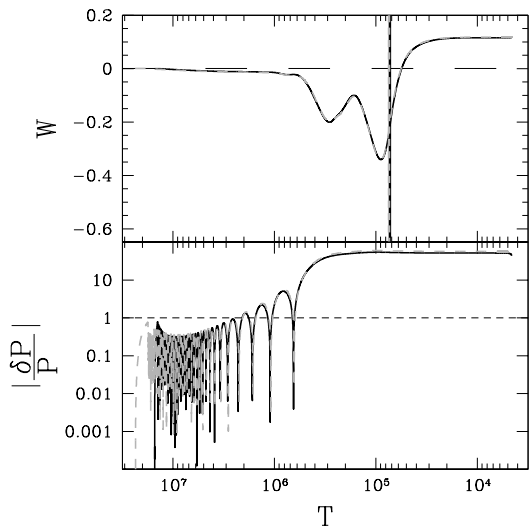}
\caption{Work integral (top panel) and eigenfunction (bottom panel) profiles for a PMS (black) and an MS (grey) modes with the same dimensionless frequency. The vertical line represents the bottom of the convective envelopes.}
\label{worksamemass}
\end{figure}

\begin{table}[!ht]
\caption{Model, angular degree, radial order, frequency and dimensionless frequency of the modes compared in the non-adiabatic section}           
\label{tablefreq}      
\centering        
\begin{tabular}{c | c c c c}     
\hline
\hline
 & $\ell$ & $k$ & frequency & dimensionless \\
 & & & (c/d) & frequency \\
\hline
 PMS $1.8 M_\odot$ & 1 & $-$43 & 0.566 & 0.188 \\
 MS $1.8 M_\odot$ & 1 & $-$51 & 0.576 & 0.188 \\
\hline                       
 PMS $2.1 M_\odot$ & 1 & $-$31 & 0.662 & 0.307 \\
 MS $1.9 M_\odot$ & 1 & $-$43 & 0.637 & 0.307 \\
\hline                       
 PMS $2.1 M_\odot$ & 1 & $-$32 & 0.637 & 0.298 \\
 MS $1.9 M_\odot$ ($\alpha = 2.07$) & 1 & $-$45 & 0.614 & 0.299 \\
\hline                       
\end{tabular}
\end{table}

\subsection{Non-adiabatic properties of PMS and MS $\gamma$~Dor}
\label{nonadiabaticproperties}

Fig.~\ref{periodsamemass} presents the period domain of  $\ell =1$ and $\ell = 2$ unstable modes for two models of $1.8 M_\odot$. One is an MS model and the other a PMS one, both with the same effective temperature $\log T_{\rm eff} =3.85$ (in the middle of $\gamma$~Dor IS, Fig.~ \ref{HRcompare} - left panel).
At this effective temperature, PMS and MS models present a similar instability domain. 
To understand the reasons for this similarity, we compare in Fig.~\ref{worksamemass} the work integrals and the eigenfunctions corresponding to two unstable modes with the same dimensionless frequency\footnote{The oscillation frequencies (in c/d) in the MS and PMS $1.8~M_\odot$ models are close but not equal because the two models have slightly different radii.} (see Table~\ref{tablefreq}). It can be seen  that for PMS and MS models the  main driving mechanism leading to the excitation of the selected modes occurs at the bottom of the CE. Both models have the same CE depth (see Sect.~\ref{internalstructure}), hence their non-adiabatic functions present the same behaviour in the external layers. The only differences between PMS and MS eigenfunctions are located in the inner layers of the models and are due to their different central structures. Nevertheless, in both cases, the amplitude of $\frac{\delta P}{P}$ in the central layers is small, and the differences between both stellar structures have no impact on the excitation of the modes.

\begin{figure}[!ht]
\centering
\includegraphics[width=70mm,height=60mm]{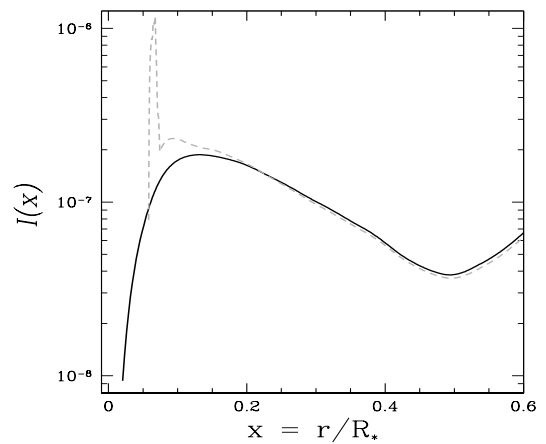}
\caption{$I(x)$ ($\propto$ radiative damping) versus the normalized radius for the $1.8 M_\odot$ PMS (full black line) and MS (dashed grey line) models.}
\label{dampingsamemass}
\end{figure}

Fig.~\ref{dampingsamemass} illustrates the variation in the term $I(x)$ (Eq.~(\ref{dimlessradiativedamping})) for the $1.8 M_\odot$ PMS and MS models. In this case, $I(x)$ (or what is the same, the radiative damping) behaves the same in both models, excepted in their very internal layers. However, the important quantity to compare is the integral of $I(x)$ on the radiative zone, $i.e.$ the area under each curve.
The close values of $\eta$ in both models explain the similarity of the period domain of unstable modes.

\begin{figure}[!ht]
\centering
\includegraphics[width=70mm,height=60mm]{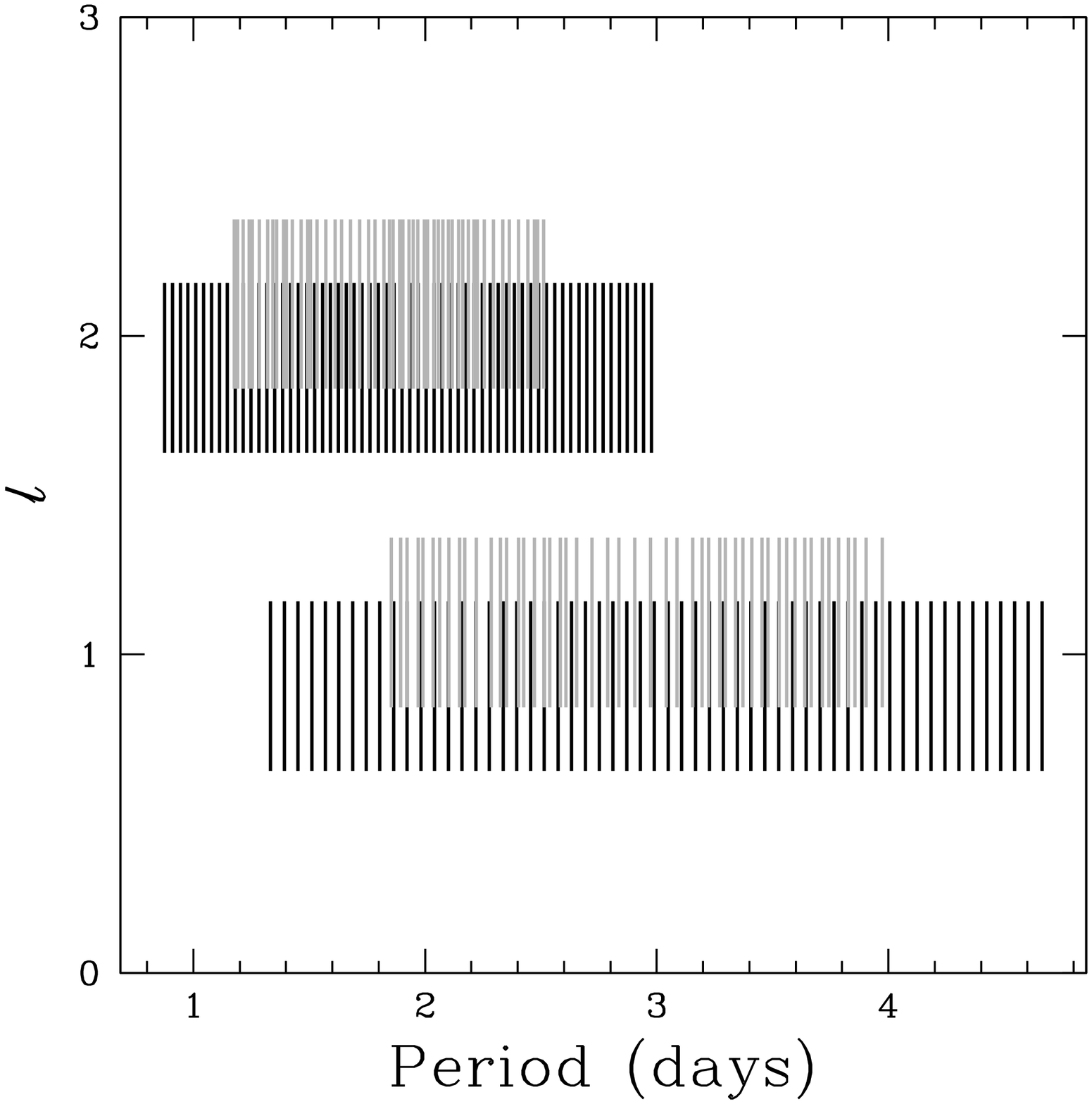}
\caption{Same figure as Fig.~\ref{periodsamemass} for the $2.1 M_\odot$ PMS and the $1.9 M_\odot$ MS models.}
\label{perioddiffmass}
\end{figure}

\begin{figure}[!ht]
\centering
\includegraphics[width=70mm,height=70mm]{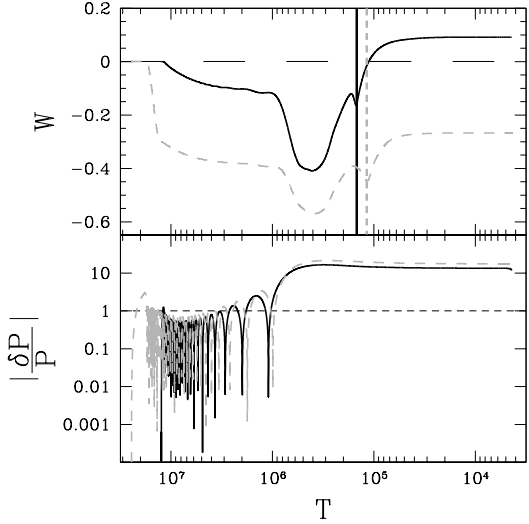}
\caption{Same figure as Fig.~\ref{worksamemass} for the $2.1 M_\odot$ PMS and the $1.9 M_\odot$ MS models.}
\label{workdiffmass}
\end{figure}

We focus now on the non-adiabatic behaviour of the $2.1 M_\odot$ PMS and the $1.9 M_\odot$ MS models, which are located close to the red border of the $\gamma$~Dor IS, where MS and PMS $g$-modes period ranges differ (see Fig.~\ref{periodteff}).
Fig.~\ref{perioddiffmass} presents the $\gamma$~Dor period ranges of these two models. While the number of unstable modes is the same for both models, the period domain corresponding to these modes is smaller for the MS model than for the PMS one. The different  $N$ profiles lead to different periods and period spacing values (see Eq.~(\ref{asymptoticperiod}) and Eq.~(\ref{meanperiodspacing})). As a consequence, the mode density is higher in the MS model than in the PMS one, and the instability range is narrower.

As for the $1.8 M_\odot$ models (Fig.~\ref{worksamemass}), we compare in Fig.~\ref{workdiffmass} the work integrals and eigenfunctions for two modes having the same dimensionless frequency. This mode is unstable in the PMS model but damped in the MS one. 
The two main differences between these models are
\begin{itemize}
\item the different driving zone depths due to different CE bottom locations, and
\item the important radiative damping that occurs in the inner layers of the MS model due to a short wavelength oscillation of the eigenfunction in the $g$-mode cavity.
\end{itemize}

\begin{figure}[!ht]
\centering
\includegraphics[width=70mm,height=60mm]{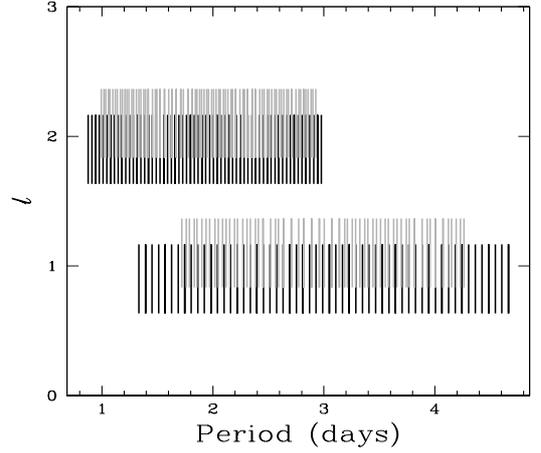}
\caption{Same figure as Fig.~\ref{periodsamemass} for the $2.1 M_\odot$ PMS and the $1.9 M_\odot$ MS models with the same temperature at the bottom of their CE.}
\label{perioddiffmassalpha}
\end{figure}

\begin{figure}[!ht]
\centering
\includegraphics[width=70mm,height=70mm]{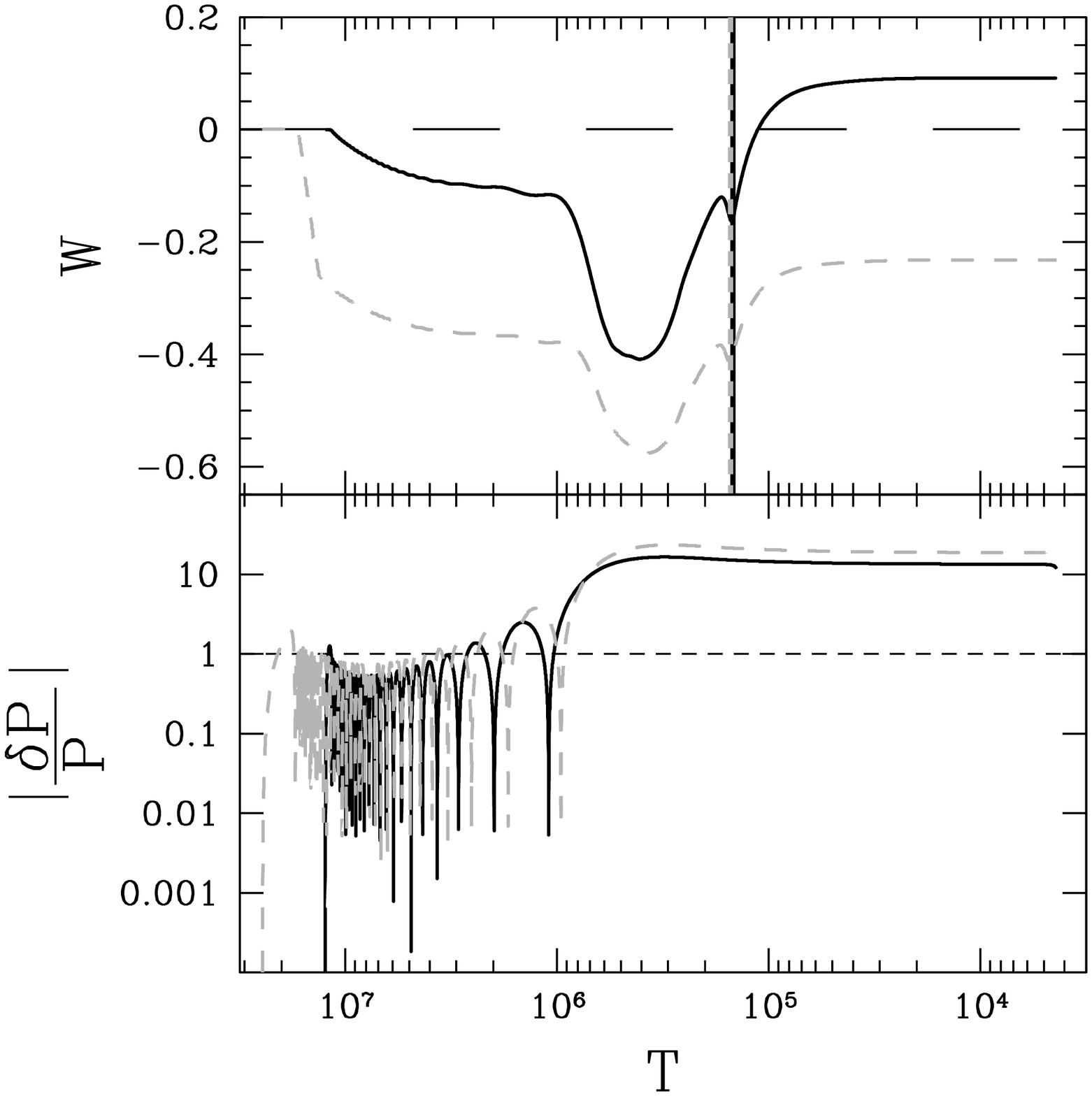}
\caption{Same figure as Fig.~\ref{worksamemass} for the $2.1 M_\odot$ PMS and the $1.9 M_\odot$ MS models with the same temperature at the bottom of their CE.}
\label{workdiffmassalpha}
\end{figure}

\begin{figure}[!ht]
\centering
\includegraphics[width=70mm,height=60mm]{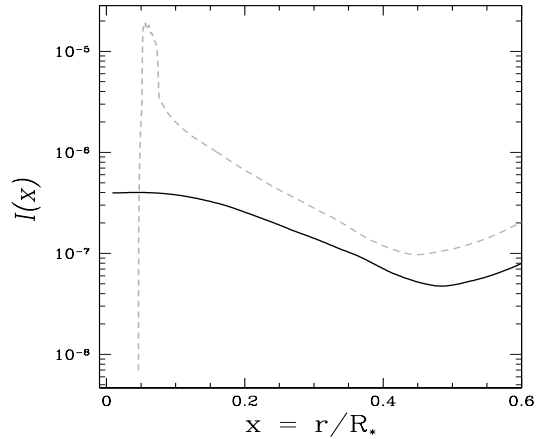}
\caption{Same figure as Fig.~\ref{dampingsamemass} for the $2.1 M_\odot$ PMS and the $1.9 M_\odot$ MS models.}
\label{dampingdiffmassalpha}
\end{figure}

To analyse the importance of each of these two factors on the damping of the MS mode, we displaced the CE bottom in the MS model by increasing the mixing length parameter from $\alpha=2.00$ to $\alpha=2.07$. We obtain a $1.9 M_\odot$ MS model with the same effective temperature and the same temperature at the base of the CE as our $2.1 M_\odot$ PMS model.
Fig.~\ref{perioddiffmassalpha} presents the same quantities and functions as Fig.~\ref{perioddiffmass} for these two models. We still obtain an MS instability range narrower than the PMS one owing the radiative damping which is much more efficient in the MS model (Fig.~\ref{workdiffmassalpha} - top panel).
In Fig.~\ref{dampingdiffmassalpha}, we present the $I(x)$ profile of the two models discussed in Fig~\ref{workdiffmass}. We clearly see that the area under the MS curve is larger than the one under the PMS profile, especially in the inner regions where the MS model presents an important $\nabla_\mu$. The corresponding sharp peak in the $N$ profile leads to short wavelength oscillations of the eigenfunctions in the central layers, hence to a strong radiative damping of the modes. The decrease in the amplitude of the eigenfunction is so important that the driving that occurs at $\log T \sim 5.3$ and at the bottom of the CE is not strong enough to drive the mode.

\section{Summary and conclusions}
\label{conclusion}

In this paper we have presented the results of a theoretical study of the seismic properties of MS and PMS models inside the observational $\gamma$~Dor instability strip. In particular, we analysed whether the stellar structure differences are reflected in adiabatic and non-adiabatic seismic features, and whether it is possible to use seismic properties to distinguish between MS and PMS evolutionary phases of $\gamma$~Dor pulsators.

The low-frequency $g$-modes that are excited in $\gamma$~Dor stars are mainly sensitive to the physical properties in the central region of the star. The more significant difference between PMS and MS models in the concerned mass domain is the presence of a chemical gradient in standard MS models, while PMS ones are almost homogeneous.  The results from the comparison of adiabatic and non-adiabatic computations for both types of models are the following

\begin{itemize}
\item Due to the evolution of the convective core in PMS and MS phases, the value of the period spacing, resulting from the first-order asymptotic approximation, changes as the stars evolve. Generally, $\langle \Delta P\rangle$  is larger for PMS models than for MS ones,  the maximum values being obtained when the convective core reaches its maximum size during the PMS phase. Nevertheless, close to the MS, PMS and MS models can have similar values of $\langle \Delta P\rangle$; therefore it will not always be possible to distinguish between both phases of evolution on the base of only the $\langle \Delta P\rangle$ value.
\item As shown by \cite{miglio2008}, the $\mu$-gradient  developed during the MS evolution leads to a periodic variation in the period spacing  whose properties are related to the location and magnitude of that gradient. On the contrary, the almost homogeneous PMS models follow the first-order asymptotic approximation quite well and present a regular pattern of $\Delta P$. Therefore, the regularity or variability of $\Delta P$ could be used as an indicator of the evolutionary state.
\item The PMS and MS instability strips overlap in the region of the HR diagram where models in both evolutionary states exist.
\item The period domain of unstable modes is quite similar for PMS and MS $\gamma$~Dor  stars except close to the red border of the IS. In this region of the HRD, evolved MS models have a huge
$\nabla_\mu$ at the CC limit, which leads to a radiative damping that is much more important than in PMS models with quasi chemically homogeneous structures, and therefore the radiative damping  affects the oscillation properties of both kinds of models very differently.
\item The non-adiabatic computations also show a regular pattern of period spacing for high-order $g$-modes in PMS models, while a periodic dependence on the radial order is obtained for $g$-modes in MS models (see Fig.~\ref{periodspacestructure}).
\end{itemize}

The comparison of these predictions with $\gamma$~Dor in young clusters will be very important for checking the current stellar structure models and the non-adiabatic theory of A-F stellar type pulsators.

\begin{acknowledgements}
We thank the anonymous referee for his/her constructive remarks.
J.M. acknowledges the Belgian Prodex-ESA for support (contract C90310).
A.M. is a Postdoctoral Researcher, Fonds de la Recherche Scientifique - FNRS, Belgium.
A.G. is supported by project PTDC/CTE-AST/098754/2008 from FCT-Portugal.
\end{acknowledgements}

\bibliographystyle{aa}
\bibliography{bouabid_16440}
\end{document}